\documentclass[11pt]{article}

\usepackage{fancyhdr}
\begin{document}

\vspace*{10mm}

\begin{center}
\uppercase{\textbf{Efficient Online Algorithmic Strategies for Several Two-Player Games with Different or Identical Player Roles}}
\end{center}

\bigskip

\begin{center}
\textsc{Mugurel Ionu\c t Andreica, Nicolae \c T\u apu\c s}
\end{center}

\bigskip

\textsc{Abstract.}
In this paper we introduce novel algorithmic strategies for effciently playing two-player games in which the players have different or identical player roles. In the case of identical roles, the players compete for the same objective (that of winning the game). The case with different player roles assumes that one of the players asks questions in order to identify a secret pattern and the other one answers them. The purpose of the first player is to ask as few questions as possible (or that the questions and their number satisfy some previously known constraints) and the purpose of the secret player is to answer the questions in a way that will maximize the number of questions asked by the first player (or in a way which forces the first player to break the constraints of the game). We consider both previously known games (or extensions of theirs) and new types of games, introduced in this paper.

\bigskip


$$
\textsc{1. Introduction }
$$

Algorithmic game theory is a topic which has been thouroughly studied because of its importance in multiple fields, like computer science, economics, social sciences or mechanism design. Game theory is used for modeling the behaviour of rational agents, both in conflicting and cooperative situations. The number of considered agents may vary from $0$, $1$ and $2$, to any number of them. Moreover, the agents may be seen as pursuing the same goal, or they may have different goals (in both cases, they may compete or collaborate).

In this paper we consider only two-player games, in most of which the two players have conflicting goals. In the first part of the paper (Sections 2-6) we discuss games in which the players have different roles. One of them has to ask questions regarding a secret pattern (e.g. tree, permutation, fake coin, and so on) and the other one has to answer the questions (usually truthfully). The first player wants to minimize the number of questions (or the asked questions and their number must satisfy some pre-established constraints), while the player who answers the questions wants to make the first player ask as many questions as possible (or force it to break the constraints of the game).

In the second part of the paper (Sections 7-8) we discuss two player games in which the players have identical roles and they compete in order to win the game. We name the considered games \emph{division games}, because their goal is to divide an initial number repeatedly at a set of given numbers, until the initial number becomes smaller than a threshold.

In Section 9 we present an extention of the well-known game in which two secret numbers are chosen, and one player is told their sum, while the other is told their product. Based on a (collaborative) conversation between the players, the original two numbers must be guessed.

For each of the games considered in Sections 2-9 we present new algorithmic strategies for playing the games (almost) optimally (given their constraints). Finally, in Section 10 we present related work and in Section 11 we conclude and discuss future work.

\lhead{}
\chead{M. I. Andreica, N. \c T\u apu\c s - Efficient Online Algorithmic Strategies for Several Two-Player Games with Different or Identical Player Roles}
\rhead{}

\begin{center}
\textsc{2. Guessing a Number with At Most One Lie}
\end{center}

Player $A$ thinks of a secret number between $1$ and $N$. Player $B$ must guess the number by asking questions of the type $Q(S)$ = is the secret number in the set $S$ ? ($S$ is a subset of $\{1,\ldots,N\}$). Player $A$ may answer with $YES$ or $NO$ and may lie at most once during the game. We would like to guess the secret number $N$ by asking as few questions as possible. We will present next a strategy which asks almost the optimal number of questions.

If player $A$ never lied, we could use binary search in order to guess the number. We would maintain an interval $[a,b]$ in which the secret number is located for sure. Initially, $a=1$ and $b=N$. While $a<b$ we:\begin{enumerate}
\item set $c=(a+b)/2$ (integer division)
\item if $Q([a,c])=YES$ then $b=c$ else $a=c+1$.
\end{enumerate}

When $a=b$, the secret number is $a$. This way, we asked $\lceil log_2(N)\rceil$ questions, which is the optimal number when no lie is allowed.

When player $A$ can lie, things get more complicated. In the first stage, we will ask $\lceil log_2(N)\rceil$ questions. Before every question $i$ ($1\leq i\leq \lceil log_2(N)\rceil$), we will have $i$ sets of numbers: $S(i,j)$ ($0\leq j\leq i-1$) is the set in which the secret number is located for sure, in case player $A$ lied at the question $j$. $S(i,0)$ corresponds to the case when player $A$ never lied (so far). Before the first question, we have $S(1,0)=[1,N]$. Let's consider the general case now, in which we are at the question $i$. Each set $S(i,j)$ is an interval $[a(i,j),b(i,j)]$. For each such set we will choose the interval $IQ(i,j)=[a(i,j), c(i,j)=(a(i,j)+b(i,j))/2]$ (where $(x+y)/2$ denotes the quotient of the integer division). The intervals $IQ(i,j)$ are disjoint, because the sets $S(i,j)$ are disjoint. Then, we will construct the set $SQ(i)$ as the union of all the intervals $IQ(i,j)$ ($0\leq j\leq i-1$) and we will ask the question $Q(SQ(i))$. If the answer is $YES$, then the new sets $S(i+1,j)$ will be equal to $[a(i+1,j)=a(i,j), b(i+1,j)=c(i,j)]$; if the answer is $NO$, the new sets $S(i+1,j)$ will be equal to $[a(i+1,j)=c(i+1,j)+1,b(i+1,j)=b(i,j)]$ ($0\leq j\leq i-1$) We will also construct the set $S(i+1,i)$, corresponding to the case when player $A$ lied at the question $i$. Thus, $S(i+1,i)=S(i,0)\setminus S(i+1,0)$. To be more precise, if the answer to the question $i$ was $YES$, then $S(i+1,i)=[a(i+1,i)=c(i,0)+1,b(i+1,i)=b(i,0)]$; otherwise, $S(i+1,i)=[a(i+1,i)=a(i,0),b(i+1,i)=c(i,0)]$.

After the first stage of the algorithm, every set $S(\lceil log_2(N)\rceil,j)$ ($0\leq j\leq \lceil log_2(N)\rceil$) will contain only one number $x(j)$. We will ask a question $Q(\{x(0)\})$. If the answer is $YES$, then $x(0)$ is the secret number. The reason is simple. If player $A$ had lied to any of the previous questions (before asking $Q(\{x(0)\})$), then $A$ would have to answer $NO$ at this question. On the other hand, if $A$ lied at the current question (but told the truth so far), then the answer should also be $NO$. If the answer to the current question is $NO$, then we know for sure that player $A$ lied once. Thus, from now on, $A$ will not be able to lie again. Thus, we will consider all the $\lceil log_2(N)\rceil +1$ numbers, $x(0),\ldots,x(\lceil log_2(N)\rceil )$ and we will perform a binary search on the set containing these numbers. We will act as if we had the interval of numbers $[0,\lceil log_2(N)\rceil ]$ at our disposal and we will use the strategy described in the beginning. The only change consists of the fact that instead of asking a question $Q([a,b])$, we will ask the question $Q(\{x(i)|a\leq i\leq b\})$. In the end, the secret number is $x(a)$.

The total number of questions is $\lceil log_2(N)\rceil+1+\lceil log_2(\lceil log_2(N)\rceil)\rceil$ (in the worst case). For instance, for $N=10^6$, we ask $26$ questions. However, the minimum number of questions for this case is $25$.

\begin{center}
\textsc{3. Guessing the Types of $M+N+1$ Persons with at Most Two Questions per Person}
\end{center}

We consider $M+N+1$ persons (numbered from $1$ to $M+N+1$), each of which is of one of the following three types: \emph{T, F, U}. Every person of type $T$ answers the truth when it is asked a question. Every person of type $F$ lies when it is asked a question. A person of type $U$ tells the truth only at every other question (i.e. tells the truth at the $1^{st}, 3^{rd}, \ldots$, question - at all the odd-numbered questions - and lies at the even-numbered questions). We can ask questions of the following type: $Q(i,j,G)$ asks the person $i$ if the person $j$ belongs to the group $G$ (where $G$ can be only $T$ or $F$); the answer to such a question is either $YES$ or $NO$. We can never have $i=j$ at a question and we can never repeat the same two persons $i$ and $j$ (in this order) as the first two arguments of a question. Moreover, we know the number of persons of each type: $M$ of type $T$, $N$ of type $F$, and one person of type $U$. We want to find the type of each person $i$ ($type(i)$) of the $M+N+1$ persons by asking questions, such that every person $i$ is asked at most $2$ questions.

We will start by considering some particular cases. If $M=N=0$ then $type(1)=U$. If $M=0$ and $N\geq 1$ then we ask each person $i$ ($1\leq i\leq N$) the question $Q(i,N+1,F)$. If $type(N+1)=U$ then all the $N$ answers will be $YES$. If $type(N+1)=F$ then we will have $N-1$ $NO$ answers and one $YES$ answer. If $N\geq 2$ then we can distinguish between the two cases:\begin{itemize}
\item if we have $N$ $YES$ answers, then all the asked persons are of type $F$ and person $N+1$ is of type $U$
\item otherwise, the only person answering $YES$ will be of type $U$ and all the others will be of type $F$
\end{itemize}

If $N=1$ then we will ask the extra question $Q(2,1,F)$. If the answer is $NO$ then $type(2)=U$ and $type(1)=F$; otherwise, $type(2)=F$ and $type(1)=U$.

The other particular case is $M\geq 1$ and $N=0$. We will ask the questions $Q(i,M+1,T)$ ($1\leq i\leq M$). If all the $M$ answers are $NO$, then $type(M+1)=U$ and the type of all the other persons is $T$. Otherwise, the answer to every question will be $YES$ (i.e. $type(M+1)=T$). If we are in this subcase and $M=1$ then we can immediately infer that $type(1)=U$. If $M>1$ then we can proceed as follows. We will ask the question $Q(M+1,1,T)$. If the answer if $NO$, then $type(1)=U$. Otherwise, we will ask the questions $Q(i,i+1,T)$ in increasing order of $i$ ($1\leq i\leq M-1$) until we obtain the first $NO$ answer. Let the answer to the question $Q(j,j+1,T)$ be $NO$. Then $type(j+1)=U$. After identifying the person of type $U$, the type of all the other persons is $T$.

We will now consider the general case, in which $M\geq 1$ and $N\geq 1$. We will start by asking the questions $Q(i,M+N+1,T)$ ($1\leq i\leq M+N$). If $type(M+N+1)=F$ (\emph{subcase 1}) then we will get $M+1$ $NO$ answers and $N-1$ $YES$ answers. Every person $i$ who answered $YES$ has type $F$. Among the $M+1$ persons who answered $NO$, $M$ are of type $T$ and one is of type $U$. If $type(M+N+1)=T$ (\emph{subcase 2}) then we will get $M$ $YES$ answers and $N$ $NO$ answers. If $type(M+N+1)=U$ (\emph{subcase 3}) then we will get $N$ $YES$ answers and $M$ $NO$ answers.

Let's consider first the semi-general case, in which $M\neq N$ and also $M\neq N-1$. In this case, we can distinguish between the three subcases we mentioned. In subcase $1$, let's assume that the $M+1$ persons who answered $NO$ are numbered $p(1),\ldots,p(M+1)$ in some arbitrary order. We will ask the question $Q(M+N+1,p(1),T)$. If the answer is $YES$ then $type(p(1))=U$. Otherwise, we will ask the questions $Q(p(i),p(i+1),T)$ in increasing order of $i$ ($1\leq i\leq M$), until we obtain the answer $NO$. Let $Q(p(j),p(j+1),T)$ be the question for which we obtained the answer $NO$. Then $type(p(j+1))=U$. After finding the person of type $U$, the type of all the other $M$ persons who answered $YES$ at the first round of questions will be $T$.

If we obtain $M$ $YES$ answers (subcase 2), then the type of each of the persons who answered $NO$ is $F$. Then, we are in the same case as when $M\geq 1$, $N=0$ (we can renumber every person who answered $YES$ with a different number from $1$ to $M$, and we can assign to the person $M+N+1$ the new number $M+1$), and the answer to each question in the first round is $YES$.

In an extension of the semi-general case where $M\neq N$ (and $M\neq N-1$ or $M=N-1$), if we obtained $N$ $YES$ answers in the first round, then:\begin{itemize}
\item every person who answered $NO$ is of type $T$
\item every person who answered $YES$ is of type $F$
\end{itemize}

A slightly more complicated case occurs when $M=N-1$ (and, obviously, $M\neq N$). In this case we cannot distinguish between subcases $1$ and $2$, because we obtain $N$ $NO$ answers and $N-1$ $YES$ answers (but subcase $3$ is distinguishable). Let's consider the $N$ persons who answered $NO$, numbered as: $p(1),\ldots,p(N)$. We will ask the questions $Q(p(i),p(i+1),T)$ ($1\leq i\leq N-1$), plus the question $Q(p(N),p(1),T)$. Let $NY$ be the number of $YES$ answers obtained at this round of questions and $NN$ be the number of $NO$ answers obtained ($NY+NN=N$). If $type(M+N+1)=T$ then the $N$ people who answered $NO$ at the first round of questions are of type $F$ and we will have $NY=N$ and $NN=0$. If $type(M+N+1)=F$ then among the $N$ people who answered $NO$ at the first round of questions are $N-1$ who are of type $T$, and one which is of type $U$. Thus, we will have $NY=N-1$ and $NN=1$ (because the person of type $U$ will lie). This way, we can identify the type of the person $M+N+1$. If $type(M+N+1)=F$ then the $N-1$ persons who answered $YES$ at the first round of questions are of type $F$, the person who answered $NO$ at the $2^{nd}$ round of questions is of type $U$ and the other $N-1$ persons (who answered $YES$ at the $2^{nd}$ round of questions) are of type $T$. If $type(M+N+1)=T$ then the $N$ people who answered $NO$ at the first round of questions are of type $F$. In order to identify the types of the $N-1$ persons who answered $YES$ at the first round of questions, we will ask the questions $Q(p(i),p(i+1),T)$ ($1\leq i\leq N-2$) and the question $Q(p(N-1),p(1),T)$ (if $N-1>1$), where $p(1),\ldots,p(N-1)$ are these $N-1$ persons. If $N-1=1$ then $p(1)$ is of type $U$. Otherwise, the only person $p(i)$ which answers $NO$ at this second round of questions is of type $U$ and the others are of type $T$.

A more complicated situation occurs when $M=N$ and we obtain an equal number of $YES$ and $NO$ answers. In this case, we cannot infer the type of the person $M+N+1$. Let's set $L=M+N+1$. Let $A$ be any person who answered $YES$ and $B$ be any person who answered $NO$ at the first round of questions. We will ask the questions $Q(L,A,T)$ and $Q(L,B,T)$ (in this order). We will denote the answers to these questions by $Ans(L,A)$ and $Ans(L,B)$, respectively. If $type(L)=T$ then $Ans(L,A)$ may be either $YES$ (if $type(A)=T$) or $NO$ (if $type(A)=U$) and $Ans(L,B)=NO$ ($type(B)=F$). If $type(L)=U$ then $Ans(L,A)=YES$ ($type(A)=F$ and the person $L$ lies) and $Ans(L,B)=YES$ ($type(B)=T$ and the person $L$ tells the truth). Thus, if $Ans(L,B)=YES$ then $type(L)=U$, the type of every person who answered $NO$ at the first round of questions is $T$, and the type of every person who answered $YES$ at the first round of questions is $F$.

If $Ans(L,B)=NO$ then $type(L)=T$. In this case, every person who answered $NO$ at the first round of questions is of type $F$. As before, we are in a case similar to the $N=0$ case. We will number all the persons who answered $YES$ at the first round of questions by $p(1),\ldots,p(M)$, such that $p(1)=A$. If $Ans(L,A)=NO$ then $type(A)=U$ and $type(p(j))=T$ ($2\leq j\leq M$). Otherwise, we will repeatedly ask the questions $Q(p(i),p(i+1),T)$ in increasing order of $i$ ($1\leq i\leq M-1$) until we obtain a $NO$ answer. Let $Q(p(j),p(j+1),T)$ be the (first) question for which the answer is $NO$. Then $type(p(j+1))=U$ and the types of all the other persons $p(i)$ ($1\leq i\leq M,i\neq j$) is $T$.

If $M=N$ or $M=N-1$ and we can uniquely identify the subcase after the first round of questions (a subcase is identified by an ordered pair of numbers, representing the number of $YES$ answers and the number of $NO$ answers at the first round of questions), then we proceed like in the semi-general case.

We should notice that in each case, every person is asked at most two questions, no person was asked a question about (him/her)self, and we never asked a question to the same person $i$ about the same person $j$. Thus, all the constraints are satisfied.

\begin{center}
\textsc{4. A Generalization of "The Counterfeit Coin" Problem}
\end{center}

We are given $n\geq 3$ coins, out of which one is different (lighter or heavier than the others). We also have a balance with two arms. We can place an equal number of coins (left or right) on each of the two arms of the balance. The balance will indicate which of the two sets of coins is heavier, or if they have the same weight. We want to identify the different coin (and whether it is lighter or heavier than the others) by using the minimum number of weightings. We will consider the more general situation, in which the results of $m$ weightings are already given and we need to minimize the number of weighting performed from now on (using the information extracted from the $m$ given weightings).

We present a dynamic programming solution, as follows. We will consider that each coin can be of one of 4 types: \emph{NM (normal)}, \emph{NH (normal or heavier)}, \emph{NL (normal or lighter)}, and \emph{NHL (normal or heavier or lighter)}. Initially, we will set the type of each coin $i$ ($1\leq i\leq N$) to be $type(i)=NHL$. Then, we will consider the $m$ given weightings. For each weighting $k$, let $L(k)$ be the set of coins located on the left arm of the balance and $R(k)$ be the set of coins located on the right arm. Let $result(k)$ be the result of the weighting $k$: $0$ if the sum of the weights of the coins in $L(k)$ and $R(k)$ are equal, $-1$ ($+1$) if the sum of the weights of the coins in $L(k)$ is smaller (larger) than that of the coins in $R(k)$. Both sets $L(k)$ and $R(k)$ contain the same number of coins. If $result(k)=0$ then we will set the type of each coin $i\in(L(k)\cup R(k))$ to $type(i)=NM$. If $result(k)=-1$ ($+1$) then:\begin{itemize}
\item at least one coin $i\in L(k)$ must have $type(i)=NHL$ or $type(i)=NL$ ($NH$), or at least one coin $i\in R(k)$ must have $type(i)=NHL$ or $type(i)=NH$ ($NL$); otherwise, the weighting is not valid (i.e. it contradicts the results of the previous weightings)
\item for each coin $i\in L(k)$:\begin{itemize}
\item if $type(i)=NHL$ then set $type(i)=NL$ ($NH$)
\item else if $type(i)=NH$ ($NL$) then set $type(i)=NM$
\end{itemize}
\item for each coin $i\in R(k)$:\begin{itemize}
\item if $type(i)=NHL$ then set $type(i)=NH$ ($NL$)
\item else if $type(i)=NL$ ($NH$) then set $type(i)=NM$
\end{itemize}
\item for each coin $i\in (\{1,\ldots,n\}\setminus (L(k)\cup R(k)))$ set $type(i)=NM$
\end{itemize}

After considering all the $m$ weightings, let $cnt_t$ be the number of coins $i$ for which $type(i)=t$ (\emph{t=NM, NHL, NH, or NL}). We have two possibilities:\begin{enumerate}
\item $cnt_{NHL}>0$ and $cnt_{NL}=cnt_{NH}=0$ (this may occur only if the result of all the $m$ weightings is $0$)
\item $cnt_{NL}\geq 0$, $cnt_{NH}\geq 0$ and $cnt_{NHL}=0$
\end{enumerate}

Note that if we have $cnt_{NHL}=cnt_{NH}=cnt_{NL}=0$ then all the coins are normal and the weightings can be considered invalid (as we assumed that exactly one counterfeit coins exists). Our dynamic programming algorithm will compute a table $nmin(n,i,j)$, where:\begin{itemize}
\item if $i\geq 0$ and $j\geq 0$ then $nmin(n,i,j)$=the minimum number of weightings which need to be performed from the state in which there are $i$ coins of type $NH$ and $j$ coins of type $NL$ (and $n-i-j$ coins of type $NM$)
\item if $j=-1$ then $nmin(n,i,-1)$=the minimum number of weightings which need to be performed from the state in which there are $i$ coins of type $NHL$ (and $n-i$ coins of type $NM$)
\end{itemize}

We will compute these values in decreasing order of the number of coins of type $NM$ belonging to a state. Let this number be $q$ ($n-1\geq q\geq 0$). Then, we have $n-q$ coins of the other types. For $q=n-1$ the computed values are: $nmin(n,1,0)=nmin(n,0,1)=0$ (as the only coin of the type $NH$ or $NL$ is the different one, and if it is of type $NH$ then it is heavier than the others, while if it is of type $NL$ it is lighter than the other coins) and $nmin(n,1,-1)=1$ (because although we know which coin is different, we need to perform an extra weighting, in order to compare its weight to that of a normal coin, in order to know if it is heavier or lighter). For $n-2\geq q$, we will procced as follows. First, we will consider all the ordered pairs $(i,j)$ such that $0\leq i$, $0\leq j$ and $i+j=n-q$. For each such ordered pair $(i,j)$, we will initialize $nmin(n,i,j)=+\infty$. Then, we will consider all the possible distinct weightings which can be performed from this state. A pseudocode close to the $C$ programming language for this case is described below:

{\it
\noindent \hspace{2mm}
{\bf for} ($a=0$; $a\leq i$ and $a\leq n/2$; $a++$)

\noindent \hspace{4mm}
{\bf for} ($b=0$; $b\leq j$ and $(a+b)\leq n/2$; $b++$)

\noindent \hspace{6mm}
{\bf if} ($a+b>0$) \{

\noindent \hspace{8mm}
{\bf for} ($c=0$; $a+c\leq i$ and $c\leq(a + b)$; $c++$)

\noindent \hspace{10mm}
{\bf for} ($d=0$; $b+d\leq j$ and $(c+d)\leq (a + b)$; $d++$)

\noindent \hspace{12mm}
{\bf if} ($(a + b) - (c + d) \leq q$) \{

\noindent \hspace{14mm}
$c_{NH}(0) = i - a - c$; $c_{NL}(0) = j - b - d$;

\noindent \hspace{14mm}
$c_{NH}(+1) = a$; $c_{NL}(+1) = d$;

\noindent \hspace{14mm}
$c_{NH}(-1) = c$; $c_{NL}(-1) = b$;

\noindent \hspace{14mm}
nmin(n,i,j)=min\{nmin(n,i,j), 1 + max\{nmin(n,$c_{NH}(0),c_{NL}(0)$),

\noindent \hspace{16mm}
nmin(n,$c_{NH}(+1),c_{NL}(+1)$), nmin(n,$c_{NH}(-1),c_{NL}(-1)$)\}\}; \}\}
}

Each weighting considers that there are $a+b$ coins placed on each arm of the balance. On the left pan there will be $a$ coins of type $NH$ and $b$ coins of type $NL$. On the right pan there will be $c$ coins of type $NH$, $d$ coins of type $NL$ and $(a+b-c-d)$ coins of type $NM$. In case the result will be $k$, then we will remain with $c_{NH}(k)$ coins of type $NH$ and $c_{NL}(k)$ coins of type $NL$.

After considering all the pairs $(i,j)$ (for the current value of $q$), we will compute the value $nmin(n,n-q,-1)$. We initialize $nmin(n,n-q,-1)=+\infty$ and then we run the following pseudocode:

{\it
\noindent \hspace{2mm}
{\bf for} ($a=0$; $a\leq n-q$ and $a\leq n/2$; $a++$)

\noindent \hspace{4mm}
{\bf for} ($b=0$; $b\leq a$ and $b\leq (n-q-a)$; $b++$)

\noindent \hspace{6mm}
{\bf if} ($a-b\leq q$) \{

\noindent \hspace{8mm}
$c_{NHL}(0) = n-q-a-b$;

\noindent \hspace{8mm}
$c_{NH}(+1) = a$; $c_{NL}(+1) = b$;

\noindent \hspace{8mm}
$c_{NH}(-1) = b$; $c_{NL}(-1) = a$;

\noindent \hspace{8mm}
nmin(n,n-q,-1)=min\{nmin(n,n-q,-1), 1 + max\{nmin(n,$c_{NHL}(0),-1$),

\noindent \hspace{10mm}
nmin(n,$c_{NH}(+1),c_{NL}(+1)$), nmin(n,$c_{NH}(-1),c_{NL}(-1)$)\}\}; \}
}

Each weighting for the case $(n-q,-1)$ considers that we place $a$ coins of type $NHL$ on the left arm. On the right arm we place $b$ coins of type $NHL$ and $a-b$ coins of type $NM$. The values $c_t(k)$ are the numbers of remaining coins of type $t$ if the result of the weighting is $k$. The time complexity of the dynamic programming algorithm is dominated by the stage of computing the values $nmin(n,i,j)$ with $i\geq 0$ and $j\geq 0$, and is of the order $O(n^6)$.

In the end, in order to solve our problem, if $cnt_{NHL}>0$ then the answer is $nmin(n,cnt_{NHL},-1)$; otherwise, the answer is $nmin(n,cnt_{NH},cnt_{NL})$. Note that a very efficient heuristic which seems correct except for some values of $n$ of a certain type, is the following. We compute the value $U=2\cdot cnt_{NHL}+cnt_{NH}+cnt_{NL}$. $U$ is the amount of uncertainty left after performing the $m$ weightings. Intuitively, since each new weighting may provide any of the $3$ possible answers, it seems plausible that there might be a weighting which reduces the uncertainty by a factor of $3$. Thus, it is plausible to assume that the number of extra questions required is around $\lceil log_3(U)\rceil$. In fact, this simple reasoning seems to provide the correct answer every time, except when $n=\frac{3^k-1}{2}$ (in these cases, the real correct answer may sometimes be larger by $1$ that the value computed by this heuristic).

\begin{center}
\textsc{5. Reconstructing a Tree by asking a Small Number of $LCA(u,v)$ Questions}
\end{center}

We consider a rooted tree with $n$ vertices. The vertices are identified with numbers from $1$ to $n$. Each vertex $i$ (except for the root of the tree, which we will denote by $r$) has a unique parent in the tree, $parent(i)$. For each vertex $i$ of the tree, we conceptually construct a list $Li(i)$ consisting of the vertex $i$ and all of its descendants:\begin{itemize}
\item if the vertex $i$ has no sons, then $Li(i)$ consists of just the vertex $i$
\item if the vertex $i$ has at least one son, then the first element of $Li(i)$ is the vertex $i$; the other elements are obtained by merging (in an arbitrary manner) the lists $Li(j)$ of the sons $j$ of the vertex $i$; when merging multiple lists $Li(j_1), \ldots, Li(j_k)$, we obtain a new list $Li'$ composed of all the elements in $Li(j_1), \ldots, Li(j_k)$ - if an element $a$ was located before an element $b$ in one of the lists $Li(j_p)$, then $a$ will also be located before $b$ in $Li'$
\end{itemize}

Given $Li(r)$, we want to reconstruct the tree. Except for knowing $Li(r)$, we may ask questions of the following type $LCA(u,v)$, which returns the lowest common ancestor in the tree of the vertices $u$ and $v$. We would like to ask as few questions $LCA(*,*)$ as possible. We will present a solution which asks $O(n\cdot log(n))$ such questions when the maximum number of sons of any vertex is upper bounded by a constant value $C\geq 2$.

We will start by presenting a simple $O(C\cdot n^2)$ solution. We will define a function $Compute(x)$ which determines the subtree rooted at $x$, given $Li(x)$. If $Li(x)=\{x\}$, then $x$ is a leaf and the function returns. Otherwise, let's assume that $x,y(1),\ldots,y(k(x))$ ($k(x)\geq 1$) are the elements from $Li(x)$ (in the order in which they occur in the list). We set $parent(y(1))=x$, we initialize $Li(y(1))=\{y(1)\}$ and we initialize $Lsons(x)=\{y(1)\}$. Then, we consider the vertices $y(j)$, in increasing order of $j$ ($2\leq j\leq k(x)$). For each vertex $y(j)$, we consider, in any order, the vertices $z$ from $Lsons(x)$. If $LCA(z,y(j))=z$ then we add $y(j)$ at the end of $Li(z)$ (and we do not consider the remaining vertices from $Li(x)$). If $LCA(z,y(j))\neq z$ for every vertex $z\in Li(x)$, then we add $y(j)$ at the end of $Lsons(x)$, we initialize $Li(y(j))=\{y(j)\}$ and we set $parent(y(j))=x$. After all this, the list $Lsons(x)$ contains all the sons $z$ of $x$ and all the lists $Li(z)$ of vertex $x$'s sons were correctly computed. Then, for every son $z$ of $x$, we call $Compute(z)$. In order to construct the tree we need to call $Compute(r)$.

Another solution is the following. Let's consider $L(1), \ldots, L(n)$, the vertices of $Li(r)$, in the order in which they appear in $Li(r)$. Obviously, $L(1)$ is the root $r$ of the tree. For every vertex $u\neq r$ of the tree we will find its parent. We have $parent(L(2))=r$. For $i=3,\ldots, n$, we will proceed as follows. The parent of the vertex $L(i)$ is one of the vertices $L(1), \ldots, L(i-1)$. We will initialize $t=r$, and then we will set all the vertices $L(2), \ldots, L(i)$ as being unmarked (the vertex $L(1)$ will be marked). While we haven't found the parent of $L(i)$, we will iterate in a loop $LP$, performing the following steps: $1)$ while $t$ has at least one unmarked son: \{$1.1)$ we will choose that son $f$ with the maximum number of vertices in its subtree; $1.2)$ we mark $f$; $1.3)$ we set $t=f$;\} $2)$ we set $t=LCA(t,L(i))$ (the lowest common ancestor of the vertices $t$ and $L(i)$); $3)$ if all the (current) sons of the vertex $t$ are marked, then we exit the loop, because $t$ is the parent of the vertex $L(i)$. In order to select each time the unmarked son $f$ with the largest number of vertices in its subtree, we will store a value $nr(u)$ for every vertex $u$, representing the number of vertices in vertex $u$'s subtree. Initially, $nr(r)=2$ and $nr(L(2))=1$. After we find the parent $t$ of every vertex $L(i)$ ($3\leq i\leq n$), we set $parent(L(i))=t$ and $nr(L(i))=1$. After this, we traverse all the ancestors $a$ of $L(i)$ (by following the $parent$ pointers starting from $parent(L(i))$ and until we reach the root $r$) and we increment $nr(a)$ by $1$.

The time complexity of the algorithm is $O(n^2)$, because, for every vertex $L(i)$, each of the vertices $L(1),\ldots,L(i-1)$ is visited (and marked) at most once. Besides, in order to obtain this time complexity, before considering a vertex $L(i)$, we need to sort the sons $f$ of each vertex $u$ in non-increasing order of $nr(f)$. Then, for every vertex $u$, we will initialize a counter $idx(u)=1$, pointing to the next unmarked son which needs to be considered (actually, it points to the index of this son in the sorted order of vertex $u$'s sons). After selecting an unmarked son $f$ (pointed to by $idx(u)$) of the vertex $u$, we will increment $idx(u)$ by $1$. If $idx(u)$ becomes greater than the number of sons of $u$, then $u$ has no more unmarked sons. Note that sorting the sons of each vertex before considering every vertex $L(i)$ is not really required. After finding $parent(L(i))$, we need to update only the lists of sons of the ancestors of $L(i)$. Let $u$ be an ancestor of $L(i)$ and let $v$ be the son of $u$ located on the path towards $L(i)$ (i.e. $parent(v)=u$). If $v=L(i)$ then we add $L(i)$ at the end of the list of sons of the vertex $u$ (as it has the smallest number of vertices in its subtree). If $v\neq L(i)$ then we need to update the position of $v$ in the list of sorted sons of the vertex $u$: we remove $v$ from this list and then we re-insert it to its appropriate position (considering the newly updated value $nr(v)$), such that the correct ordering of vertex $u$'s sons is maintained.

Let's now analyze the number of questions $LCA(a,b)$ asked by the algorithm. Let's notice that we only ask one question for the whole group of vertices marked during step $1$ of one iteration of the $LP$ loop. Let's consider the vertex $t$ obtained at the end of an iteration of the $LP$ loop. If $t$ is not the parent of $L(i)$, then at the next iteration of the $LP$ loop we will not consider any vertex in the subtrees of the marked sons $f$ of the vertex $t$ which were selected during the current or previous iterations. Thus, the number of vertices which are still potential parents for $L(i)$ is at most $(C-nsel(t))\cdot nr(t)/C$, where $nsel(t)$ denotes the number of marked sons of the vertex $t$. Thus, after at most $C-1$ consecutive questions, the number of vertices which are potential parents of $L(i)$ drops by a factor of $C$. This proves that the total number of questions asked for finding $parent(L(i))$ is $O((C-1)\cdot log_C(n))=O(log(n))$. The total number of questions is $O(n\cdot (C-1)\cdot log_C(n))=O(n\cdot log(n))$.

\begin{center}
\textsc{6. Reconstructing a Permutation by Asking a Bounded Number of Distance Questions per Element}
\end{center}

We consider an unknown permutation with $N$ elements (numbered from $1$ to $N$). We want to reconstruct the permutation by asking a small number of questions of the following type: $D(i,j)$ asks for the distance between the elements $i$ and $j$ (i.e. the absolute difference between their positions in the permutation). In fact, we would like for each element $i$ to occur at most $3$ times as an argument to a question $D(i,j)$ (or $D(j,i)$).

We will assume that element $1$ is located on position $0$ and we will determine the positions of the other elements relative to this. We will denote by $x(i)$ the position of the element $i$. We will start by asking the questions $D(1,2)$, $D(1,3)$ and $D(2,3)$. From these questions we will be able to compute exactly the positions of the elements $2$ and $3$. For instance, if $D(1,2)+D(1,3)=D(2,3)$ we will have $x(2)=-D(1,2)$ and $x(3)=D(1,3)$; if $D(1,2)+D(2,3)=D(1,3)$ then $x(2)=D(1,2)$ and $x(3)=D(1,3)$, and so on.

We will now consider the elements $4,\ldots,N$ (in this order), in pairs of two conscutive elements. Let's assume that we are now considering the elements $i$ and $i+1$ and that the positions of all the elements $1,\ldots,i-1$ have already been computed. During the algorithm we will maintain the following invariant. We will always have $3$ elements from the set $\{1,\ldots,i-1\}$ which occurred only two times as an argument to a question. Let these elements be $x$, $y$, and $z$. Initially, $x=1$, $y=2$ and $z=3$. When considering the elements $i$ and $i+1$ we will first ask the question $D(i,i+1)$. Then, we will choose two elements $a$ and $b$ from the set $\{x,y,z\}$ and we will ask the questions $D(a,i)$ and $D(b,i+1)$. Let's now analyze the problem locally. We have $4$ elements: $a$, $b$, $i$ and $i+1$. We know $x(a)$, $x(b)$, and the distances $D(i,i+1)$, $D(a,i)$ and $D(b,i+1)$. The only two possibilities for $x(i)$ are $x(a)-D(a,i)$ and $x(a)+D(a,i)$, while for $x(i+1)$ are $x(b)-D(b,i+1)$ and $x(b)+D(b,i+1)$. We will consider all the $2$x$2$ possibilities and we would like to have only one valid possibility, i.e. only one possibility for which $|x(i)-x(i+1)=D(i,i+1)$. In most cases, the solution will be unique. However, if $Dab=|x(b)-x(a)|=D(i,i+1)$ and $D(a,i)=D(b,i+1)$, then there are two valid solutions among the 4 possibilities. We could solve the ambiguity by asking the question $D(a,i+1)$ (which will only be satisfied by one of the two valid solutions), but this would mean that element $a$ was given as an argument to $4$ questions. We can avoid this case by carefully choosing the elements $a$ and $b$ from the set $\{x,y,z\}$. We will choose two elements $a$ and $b$ such that $|x(b)-x(a)|\neq D(i,i+1)$. Note that there are three distances between the three elements $x$, $y$, and $z$, out of which at most $2$ can be equal. Thus, we can always find a pair $(a,b)$ among $x$, $y$, and $z$, such that $|x(b)-x(a)|\neq D(i,i+1)$.

After finding the positions $x(i)$ and $x(i+1)$, the new set of three elements which were given as question arguments only two times will be $\{i,i+1\}\cup (\{x,y,z\}\setminus \{a,b\})$.

If $N$ is even, in the end, we will have an element ($N$) which has no pair. For this element we will choose two elements $a$ and $b$ among $x$, $y$ and $z$, and we will ask the questions $D(a,N)$ and $D(b,N)$. Based on these distances and on $Dab=|x(a)-x(b)|$ we will uniquely determine $x(N)$.

After all these we just need to shift the positions of the elements to the interval $[1,N]$. We will compute $xmin=min\{x(i)|1\leq i\leq N\}$. Then, we will modify each position as follows: $x(i)=x(i)-xmin+1$ ($1\leq i\leq N$).

\begin{center}
\textsc{7. The Division Game with Integer Numbers}
\end{center}

We consider a natural number $N\geq 0$ and a list of $K\geq 1$ distinct natural numbers: $P(1),\ldots, P(K)$ ($P(i)\geq 2$; $1\leq i\leq K$). Two players take turns alternately. At its turn, a player will replace the number $N$ by any number $\lfloor\frac{N}{P(i)}\rfloor$, where $1\leq i\leq K$. If $N\leq L$ (for a given $L\geq 0$) then the player who has to perform the next move loses the game. A simple dynamic programming strategy is the following. For every natural number $q$ ($0\leq q\leq N$) we compute $win(q)=1$ if the current player has a winning strategy when its current value is $q$, or $0$, otherwise. $win(0\leq q\leq L)=0$. For $L+1\leq q\leq N$ we consider all the numbers $q'=\lfloor\frac{q}{P(i)}\rfloor$ ($1\leq i\leq K$). If we have $win(q')=0$ for at least one such number, then $win(q)=1$; otherwise (if $win(q')=1$ for all the values $q'$) then $win(q)=0$.

The problem with this approach is that it is inefficient for large values of $N$, because it has to compute $O(N)$ $win(*)$ values. We will use a recursive approach instead, coupled with memoization. We will maintain a hash table $H$ with pairs $(key=q, value=win(q))$. Then, we will call the function $computeWin(N)$. $computeWin(q)$ computes the value $win(q)$ and returns it. The function works as follows. If the key $q$ is located within $H$, then it returns the value associated to the key $q$. Otherwise, if $q\leq L$, it returns the value $0$. If $q\geq L+1$, the function considers every value $q'=\lfloor\frac{q}{P(i)}\rfloor$ ($1\leq P\leq K$). For each such value, it performs the call $computeWin(q')$ in order to obtain the value $win(q')$. Then, $win(q)$ is computed using the same rules as before. After computing $win(q)$, the pair $(key=q, value=win(q))$ is inserted into $H$ and the value $win(q)$ is returned. The number of processed values $q$ is significantly smaller than $N$.

A much simpler solution when the list of numbers $P(1),\ldots,P(K)$ is $2,\ldots,$ $K+1$ and $L=0$ is the following. If $N=L$ then the first player to move loses the game. Otherwise, we set $Q=N$ and $M=K+1$. While $Q\geq(2\cdot M)$ we set $Q=Q$ \emph{div} $(2\cdot M)$ (integer division). In the end, if $Q<M$ the first player to move (for the number $N$) has a winning strategy; if $Q\geq M$ then the second player to move has a winning strategy.

\begin{center}
\textsc{8. The Division Game with Real Numbers}
\end{center}

In this section we consider the same game as in the previous section, except that the division is a real division (not integer), and the numbers $N$, $L$ and $P(i)$ ($1\leq i\leq K$) are real numbers (moreover, $P(i)>1.0$). We can use the same solution based on memoization as in the previous problem, but now the number of distinct numbers encountered would be too large. In this case, we will divide the real axis into disjoint intervals of equivalent numbers. The equivalence of two numbers $a$ and $b$ implies, among other things, that the (optimal) result of the game when $N=a$ is the same as when $N=b$. For each interval $(a,b]$ of equivalent numbers we will compute its winning value: $win((a,b])=1$, if the next player to move wins the game when the current value is a number from the interval $(a,b]$, and $0$, otherwise. The initial interval is $(0,L]$ and $win((0,L])=0$.

A first solution is the following. We will maintain a balanced tree $T$ with the intervals computed so far and a heap $H$ (initially empty). Let $(a,b]$ be the last interval computed. We will compute the intervals from left to right. Initially, $a=0$, $b=L$ and $T$ contains only the interval $(0,L]$. While $b<N$ we perform the following steps. We will insert into $H$ the values $b\cdot P(i)$ ($1\leq i\leq K$). Then, we extract from $H$ the minimum value $x$. We set $a=b$ and then $b=x$. We will compute $win((a,b])$ as follows. We choose a number $y$ from the interval $(a,b]$ (e.g. $y=(a+b)/2$ or $y=b$). Then, we consider all the values $y'=y/P(i)$ ($1\leq i\leq K$). For each value $y'$, we search $T$ (in $O(log(|T|))$ time) in order to find the interval $(u,v]$ containing $y'$. If $win((u,v])=0$ then we set $win((a,b])=1$. If none of the values $win((u,v])$ is $0$ (for all the values $y'$), then we set $win((a,b])=0$. Afterwards, we insert the interval $(a,b]$ into $T$. As soon as $b\geq N$ we stop. The result of the game is determined by the value $win((a,b])$. The disadvantage of this approach is that it may end up computing many intervals. Moreover, there may be many consecutive intervals with the same $win$ value. Collapsing all such intervals into one larger interval will help from the memory point of view, but not from that of the running time.

A more efficient approach is presented next. First, we will consider a data structure $DS$ which will store disjoint intervals $(a,b]$ and which supports the following types of operations:\begin{itemize}
\item insert an interval $(a,b]$ into $DS$\begin{itemize}
\item before inserting it, all the intervals $(c,d]$ fully included in $(a,b]$ are removed from $DS$
\item if there is an interval $(c,d]$ with $c<a$ and $a\leq d\leq b$ in $DS$ then we set $a=c$ and then we remove $(c,d]$ from $DS$
\item if there is an interval $(c,d]$ with $d>b$ and $a\leq c\leq b$ in $DS$ then we set $b=d$ and then we remove $(c,d]$ from $DS$
\item if there is no interval $(c,d]$ with $c\leq a$ and $d\geq b$ in $DS$ then we insert $(a,b]$ into $DS$
\end{itemize}
\item find the interval $(a,b]$ with the minimum value of $a$
\item remove a given interval $(a,b]$ from $DS$
\item find the interval $(a,b]$ with the smallest value $b$ such that $b\geq x$ for a given value of $x$
\end{itemize}

$DS$ can be implemented easily by using any balanced tree. Every operation is supported in $O(log(m))$ time (where $m$ is the number of intervals in $DS$), except for the insertion operation, which takes $O((q+1)\cdot log(m))$ time (or even $O(q+log(m))$ time), where $q$ is the number of intervals intersected by the newly inserted interval $(a,b]$.

We will maintain $K+1$ data structures $DS$. The first one, $T_1$ will contain intervals for which the first player to move will be the winner (considering an optimal strategy). Then, we will have $K$ data structures $T_0(i)$ ($1\leq i\leq K$), with the property that for any number $x$ of an interval $(a,b]$ from $T_0(i)$, $x/P(i)$ belongs to an interval $(c,d]$ with $win(c,d]=1$. We consider the operation of \emph{expanding} an interval $(a,b]$. For each value $P(i)$, we compute an interval $(c(i)=max\{P(i)\cdot a,b\},d(i)=P(i)\cdot b]$, such that for every $x\in (c(i),d(i)]$, we have that $x/P(i)\in (a,b]$ ($1\leq i\leq K$). Then, we insert every such interval $(c(i),d(i)]$ in $T_1$. We also consider the operation of $K-expanding$ an interval $(a,b]$. For each value $P(i)$ we compute the same interval $(c(i),d(i)]$ mentioned earlier and we insert it into $T_0(i)$ ($1\leq i\leq K$).

The algorithm proceeds as follows. Let $(a,b]$ be the last computed interval. Initially, we have $a=0$, $b=L$, and $win((a,b])=0$. While $b<N$ we proceed as follows:\begin{enumerate}
\item for each data structure $T_0(i)$ ($1\leq i\leq K$), while $T_0(i)$ is not empty and the interval $(c(i),d(i)]$ with the minimum value of $c(i)$ has the property that $d(i)\leq b$, we remove $(c(i),d(i)]$ from $T_0(i)$.
\item if $T_1$ is not empty, then let $(c,d]$ be the interval with the minimum value of $c$ in $T_1$: if $c=b$ then:\begin{enumerate}
\item if $win((a,b])=1$ then set $b=d$ else set $(a,b]=(c,d]$
\item set $win((a,b])=1$
\item \emph{K-expand} the interval $(a,b]$
\item remove the interval $(c,d]$ from $T_1$
\end{enumerate}

\item if, however, $T_1$ is empty or $c>b$ then:\begin{enumerate}
\item let $(u(i),v(i)]$ be the interval with the smallest value $v(i)$ from $T_0(i)$ such that $v(i)\geq b$ ($1\leq i\leq K$)
\item let $bmin=min\{b(i)|1\leq i\leq K\}$
\item if $win((a,b])=0$ then set $b=bmin$ else set $a=b$ and then $b=bmin$
\item set $win((a,b])=0$
\item \emph{expand} the interval $(a,b]$
\end{enumerate}
\end{enumerate}

In the end, if $win((a,b])=1$ (with $a<N\leq b$) then the first player to move wins; otherwise, the second player has a winning strategy. The time complexity is $O(M\cdot K\cdot log(M))$, where $m$ is the total number of iterations of the \emph{"while ($b<N$)"} loop.

\begin{center}
\textsc{9. An Extension of the Sum-Product Game}
\end{center}

We consider an extension of the well-known game concerning the sum and product of two numbers. There are two players, $S$ and $P$. $S$ knows the sum of two numbers $a$ and $b$, while $P$ knows their product. $a$ and $b$ are integer numbers from the interval $[1,N]$. We consider both the case when $a$ and $b$ must be distinct numbers, as well as the case when they may be equal. A conversation between $S$ and $P$ takes place. Each player makes an affirmation alternately. There are $M+1$ affirmations made overall by the two players. Each of the first $m$ affirmations is "I don't know the numbers.". The last affirmation is "I know the numbers.". We would like to know all the possible pairs of numbers $(a,b)$ which could have generated the given conversation.

We will start by generating all the sums (products) of all the possible pairs $(a,b)$. For each sum $s$ (product $p$) we will store the number of valid pairs $(a,b)$ whose sum (product) is $s$ ($p$): $snum(s)$ ($pnum(p)$). We can compute this in $O(N^2)$ time, using a hash table $HS$ ($HP$) in which the key is the sum (product) and the value is the number of pairs encountered so far (while generating all the valid pairs) whose sum (product) is $s$ ($p$). Moreover, we will also maintain a set $SP$ with all the possible valid pairs (initially, $SP$ contains all the valid pairs $(a,b)$). $SP$ can be implemented as a hash table, too. Then, we will perform $m$ rounds of eliminating pairs. Let's assume that we are at round $R$ ($1\leq R\leq M$). If $R$ is odd, it is the turn of the player $S$ to make an affirmation; otherwise, it is the turn of player $P$. If it $S$'s ($P$'s) turn, we will consider all the sums (products) $s$ ($p$) with $snum(s)=1$ ($pnum(p)=1$). For each such sum (product) $s$ ($p$), we will remove from $SP$ the pair $(a,b)$ with $a+b=s$ ($a\cdot b=p$). After removing $(a,b)$ from $SP$, we will also decrease by $1$ the values $snum(a+b)$ and $pnum(a\cdot b)$. After performing the $m$ rounds, if $M+1$ is odd (even), then the last affirmation is made by $S$ ($P$). If it is $S$'s ($P$'s) turn, then we will report as possible solutions all the pairs $(a,b)$ with $snum(a+b)=1$ ($pnum(a\cdot b)=1$).

\begin{center}
\textsc{10. Related Work}
\end{center}

Guessing secret numbers when lies are allowed has been previously considered in several papers, like [5] and [6]. Our solution, however, is new and of independent theoretical interest.

The counterfeit coin problem (without the initial given weightings) has been considered in many papers (e.g. [1, 2, 3, 4]), from multiple perspectives, like multiple counterfeit coins, having the knowledge that the different coin is lighter (heavier) than the others, obtaining mathematical equations for the minimum number of required weightings, and so on. [4] presents a greedy algorithm for the counterfeit coin problem, focused on reducing the uncertainty as much as possible. Although that algorithm starts from the case when no weightings are given, it is straight-forward to run that algorithm from the state obtained after considering all the initial given weightings.

Algorithms for reconstructing trees efficiently have been considered in many papers (e.g. [7, 8]), because of their applications in biology (reconstruction of philogeny trees).

(Multi-)permutation guessing problems using different types of questions were considered in several papers. In [9], the problem of guessing a permutation by asking questions in which the argument is a candidate permutation and the answer is the number of positions in which the secret permutation and the candidate permutation coincide. In [10], a similar problem was considered, but for multi-permutations (with known number of occurrences of each element).

A reference book in algorithmic game theory is [11], in which many topics regarding both collaborative and conflicting agents are considered. However, the types of situations considered in [11] are of a somewhat different nature than the ones considered in this paper.

Besides the published material, we are aware of several related problems whose solutions were mentioned to us in personal communications. We will briefly discuss some of these problems and their solutions here, with the permission of the solutions' authors.

The first problem considers the reconstruction of a tree from distance data between leaves. We know the number $K\geq 3$ of leaves of a tree. The leaves are numbered from $1$ to $K$. We can ask questions of the type $D(x,y)$ for which the answer is the number of edges on the unique path between the vertices $x$ and $y$ in the tree (both $x$ and $y$ must be leaves). An $O(K^2)$ algorithm is presented first. We ask for the distance $D(1,2)$ between the leaves $1$ and $2$ of the tree. Then we ask the distances $D(1,3)$ and $D(2,3)$. Based on this information we can find at which vertex $x$ on the path from $1$ to $2$ branches the path towards the leaf $3$. Let $D(i,j)$ be the distance between the vertices $i$ and $j$ (whether they are leaves or not). We have $D(1,x)+D(x,3)=D(1,3)$ and $D(1,2)-D(1,x)+D(x,3)=D(2,3)$. From this system with two equations and two unknown variables we can easily compute $D(1,x)$ and $D(x,3)$. We will add a path from $x$ to $3$, containing $D(x,3)-1$ internal vertices. For each internal vertex $p$ we will maintain a list of leaves $L(p)$ (which contains all the leaves $q$ for which a path from $p$ to $q$ was added in the tree). At the moment when an internal vertex $p$ is added to the tree, the list $L(p)$ will be empty. Initially, only the list $L(x)$ ($x$ was computed as described above) will contain the leaf $3$. $|L(p)|$ will denote the number of elements in the list $L(p)$ (this number is maintained as a counter which is incremented by $1$ every time a new element is added to the list). We will also maintain an index $idx(x)$ and a timestamp $tstamp(x)$ for each internal vertex $x$ (when the vertex is added to the tree, these values are initialized to $0$).

We denote by $B(u,v,q)$=the vertex at which the path between $u$ and $q$ branches from the path between $u$ and $v$. We will store all the computed values $B(u,v,q)$. We saw earlier how we can compute $B(u,v,q)$ when we know the distances $D(u,v)$, $D(u,q)$ and $D(v,q)$ (just substitute $1$, $2$ and $3$ by $u$, $v$ and $q$ in the previous paragraph). Then, for each leaf $i=4,\ldots,K$, we proceed as follows. We start with $a=1$ and $b=2$. While $idx(B(a,b,i))<|L(B(a,b,i))|$:\begin{enumerate}
\item let $x=B(a,b,i)$
\item let $j$ be the leaf on the position $idx(x)+1$ from the list $L(x)$
\item set $idx(x)=idx(x)+1$
\item set $b=j$
\end{enumerate}

During the algorithm we never ask the same question twice (i.e. we store the values $D(i,j)$ in a hash table after they are asked for). When computing $B(a,b,q)$ we ask all the distances which are still unknown (among $D(a,b)$, $D(a,q)$ and $D(b,q)$). In the end, let $x=B(a,b,i)$. We add a path from the vertex $x$ to the leaf $i$, containing $D(x,i)-1$ internal vertices. Note that this distance is computed as part of finding $B(a,b,i)$. Then we add the leaf $i$ at the end of $L(x)$. The presented algorithm may ask $O(K^2)$ questions.

An improvement which asks $O(K\cdot log(K))$ questions is the following. We will determine an order $o(1),\ldots,o(K)$ of the $K$ leaves, such that they correspond to the order in which they may be visisted by a DFS traversal starting from the vertex adjacent to $o(1)$. The property of this ordering is that $D(o(i),B(o(i),o(j),o(q)))\leq D(o(i),B(o(i),o(j),o(q'))$, for $i<q<q'<j$. After computing this order, the tree can be constructed easily. We first add the path between $o(1)$ and $o(2)$. Then, for $3\leq i\leq K$ (in this order), we add the path from $B(o(i-2),o(i-1),o(i))$ to $o(i)$. In order to find this ordering we will start with an initial ordering consisting of the leaves $o(i)=i$ ($1\leq i\leq 3$). Then we will consider every leaf $4\leq i\leq K$. We will binary search the position where this leaf will be inserted. We will start with an interval $[a=1,b=i-1]$. While $b-a\geq 2$ we perform the following steps:\begin{enumerate}
\item \emph{c=(a+b) div 2} (integer division)
\item let $x=B(o(a),o(b),i)$ and $y=B(o(a),o(b),c)$
\item if $D(o(a),x)\leq D(o(a),y)$ then $b=c$ else $a=c$
\end{enumerate}

In the end, we insert the leaf $i$ between the positions $a$ and $b=a+1$ in the leaf ordering. It is obvious that only $O(log(K))$ questions are asked for each leaf. The $O(K^2)$ solution is an original solution for this problem, but the idea for the $O(K\cdot log(K))$ refinement was mentioned to us by A. Vladu in a personal communication.

The second problem, whose solution was mentioned to us by N. Mo\c t in a personal communication, is the following. There is a secret (ordered) tuple $(x,y,z)$, where $x$, $y$ and $z$ are (not necessarily distinct) numbers from the set $\{1,\ldots,N\}$. In order to find the secret tuple, we can ask questions of the following type: $Ask(a,b,c)$. The answer to a question $Ask(a,b,c)$ is $1$ if at least two values from the multiset $\{x-a,y-b,z-c\}$ are zero, and $0$ otherwise. We want to find the secret tuple $(x,y,z)$ using as few questions as possible.

We will denote by $try(a,b,c)$ the answer to the question $Ask(a,b,c)$. Using at most $N^2$ questions, we can find a tuple $(a,b,c)$ (in which not all the three numbers are equal), such that $try(a,b,c)=1$. For this, we will consider every possible values for $a$ and $b$, while $c$ will be chosen such that it is different from both $a$ and $b$ (if such a value exists). Note that finding such a tuple is always possible for $N\geq 2$ (and the problem is trivial for $N=1$). Then, by using $3$ more questions, we will identify which of the three positions coincide with the positions from the secret tuple. We swap, one at a time, $a$ and $b$ (obtaining the tuple $(b,a,c)$), then $a$ and $c$ (obtaining the tuple $(c,b,a)$), and then $b$ and $c$ (obtaining the tuple $(a,c,b)$). The pair $(u,v)$ of swapped numbers for which $try(e,f,g)=0$ (where $(e,f,g)$ is the tuple obtained after the swap) determines the two positions which coincide with the secret tuple (i.e. the numbers $u$ and $v$ coincide with the numbers on the same positions as $u$ and $v$ from the secret tuple). Then, using $N$ more questions, we can also identify the $3^{rd}$ number. We will replace the number $u$ by a value $u'$ different from $u$ in the tuple $(a,b,c)$. Then, we will consider each of the $N$ possible values for the $3^{rd}$ number (the one different from $u$ and $v$). The value $w$ for which we get a $1$ answer for the obtained tuple is the correct value for the $3^{rd}$ number. Thus, we were able to find the secret tuple using at most $N^2+N+3$ questions.

However, we can do better than this. In particular, the initial stage of finding a tuple in which two numbers coincide with the corresponding numbers from the secret tuple can be optimized. We will split the set $\{1,\ldots,N\}$ into two sets $S_1$ with $\lfloor N/2 \rfloor$ elements and $S_2$ with $N-\lfloor N/2 \rfloor$ elements, respectively. This way we can be certain that at least two of the numbers from the secret tuple belong to the same set (either $S_1$ or $S_2$). Then, with at most $\lfloor N/2\rfloor \cdot \lfloor N/2\rfloor$ + $(N-\lfloor N/2\rfloor) \cdot (N-\lfloor N/2\rfloor)$ questions we can find a "good" tuple (for which two of his numbers coincide with the corresponding numbers from the secret tuple). Let the numbers of the set $S_i$ be numbered as $x(i,0),\ldots,x(i,k(i)-1)$ ($i=1,2$). For each set $i$ ($i=1,2$) we will ask questions for tuples of the form $(x(i,u),x(i,v),x(i,w))$ such that $(u+v+w)$ \emph{mod} $k(i)=0$. After choosing the indices $u$ and $v$ for such a tuple, we can choose the index $w$ in only a single way (it is uniquely determined by $u$ and $v$: $w=(2\cdot k(i)-u-v)$ \emph{mod} $k(i)$). We ask $k(i)^2$ questions for each set $i$ ($i=1,2$), where $k(i)$ is the cardinality of the set $i$. It is guaranteed that we will find a tuple $(a,b,c)$ for which $try(a,b,c)=1$ among the considered tuples.

\begin{center}
\textsc{11. Conclusions and Future Work}
\end{center}

In this paper we presented novel algorithmic strategies for playing optimally several two player games, in which the players may have different or identical roles. The considered games are either previously studied games (or extensions of theirs), or new games which are introduced in this paper. The games and the presented strategies fit quite nicely in the framework provided by the domain of algorithmic game theory. All the discussed solutions are specific to each problem. As future work, we intend to research the possibility of devising a more generic algorithmic framework for computing optimal strategies for multiple two-player games with different player roles, similar to the ones mentioned in this paper.

\begin{center}
\textsc{Acknowledgements}
\end{center}

The work presented in this paper has been supported by CNCSIS-UEFISCSU under research grants PD\_240/2010 (contract no. 33/28.07.2010) and ID\_1679/2008 (contract no. 736/2009), and by the Sectoral Operational Programme Human Resources Development 2007-2013 of the Romanian Ministry of Labour, Family and Social Protection through the Financial Agreement POSDRU/89/1.5/S/62557.

\bigskip
$$
\textsc{References}
$$

[1] M. Aigner, and L. Anping, Searching for Counterfeit Coins, Graphs and Combinatorics, vol. 13, (1997), pp. 9-20.

[2] B. Manvel, Counterfeit Coin Problems, Mathematics Magazine, vol. 50, no. 2, (1977), pp. 90-92.

[3] L. Pyber, How to Find Many Counterfeit Coins?, Graphs and Combinatorics, vol. 2, (1986), pp. 173-177.

[4] M. I. Andreica, Algorithmic Decision Optimization Techniques for Multiple Types of Agents with Contrasting Interests, Metalurgia International, vol. 14, special issue no. 11, (2009), pp. 162-170.

[5] J. Spencer, Guess a Number - with Lying, Mathematics Magazine, vol. 57, no. 2, (1984), pp. 105-108.

[6] F. Cicalese, and U. Vaccaro, Optimal Strategies against a Liar, Theoretical Computer Science, vol. 230, (1999), pp. 167-193.

[7] J. J. Hein, An Optimal Algorithm to Reconstruct Trees from Additive Distance Data, Bulletin of Mathematical Biology, vol. 51, no. 5, (1989), pp. 597-603.

[8] V. King, L. Zhang, and Y. Zhou, On the Complexity of Distance-based Evolutionary Tree Reconstruction, Proceedings of the $14^{th}$ ACM-SIAM Symposium on Discrete Algorithms, (2003), pp. 444-453.

[9] K.-I. Ko, and S.-C. Teng, On the Number of Queries Necessary to Identify a Permutation, J. of Algorithms, vol. 7 (4), (1986), pp. 449-462.

[10] M. I. Andreica, A. Grigorean, N. Tapus, Algorithms for Identifying Sequence Patterns with Several Types of Occurrence Constraints, Proc. of the IEEE Intl. Conf. on Symbolic and Numeric Algorithms (SYNASC), (2009).

[11] N. Nisam, et al., Algorithmic Game Theory, Cambridge University Press, 2007.

\bigskip

\noindent 
Mugurel Ionu\c t Andreica, Nicolae \c T\u apu\c s\\
Department of Computer Science and Engineering\\
Politehnica University of Bucharest\\
Splaiul Independen\c tei 313, sector 6, Bucharest, Romania\\
email:\textit{\{mugurel.andreica, nicolae.tapus\}@cs.pub.ro}
\end{document}